\documentclass[superscriptaddress, twocolumn, prl]{revtex4}
\usepackage{graphicx}   % need for figures

\newcommand{\eqnref}[1]{(\ref{#1})}

\newcommand{\figref}[1]{\figurename~\ref{#1}}

\newcommand{\meter}[1][]{\ifx|#1|\unit{m}\else\unit[#1]{m}\fi}
\newcommand{\hertz}[1][]{\ifx|#1|\unit{Hz}\else\unit[#1]{Hz}\fi}
\newcommand{\fm}[1][]{\ifx|#1|\unit{fm}\else\unit[#1]{fm}\fi}
\newcommand{\fluence}[1][]{\ifx|#1|$\unit{mJ/cm^2}$\else$\unit[#1]{mJ/cm^2}$\fi}
\begin{document}

\title{Structural and magnetic dynamics of a laser induced phase
transition in FeRh}

\author{S. O. Mariager}
\email{simon.mariager@psi.ch}
\affiliation{Swiss Light Source, Paul
Scherrer Institut, 5232 Villigen, Switzerland}
\author{F. Pressacco}
\affiliation{Fakult\"{a}t f\"{u}r Physik, University of Regensburg,
93053 Regensburg, Germany}
\author{G. Ingold}
\affiliation{Swiss Light Source, Paul Scherrer Institut, 5232
Villigen, Switzerland}
\author{A. Caviezel}
\affiliation{Swiss Light Source, Paul Scherrer Institut, 5232
Villigen, Switzerland}
\author{E. M\"{o}hr-Vorobeva}
\affiliation{Swiss Light Source, Paul Scherrer Institut, 5232
Villigen, Switzerland}
\author{P. Beaud}
\affiliation{Swiss Light Source, Paul Scherrer Institut, 5232
Villigen, Switzerland}
\author{S. L. Johnson}
\affiliation{Swiss Light Source, Paul Scherrer Institut, 5232
Villigen, Switzerland}
\author{C. J. Milne}
\affiliation{\'{E}cole Polytechnique Fed Lausanne, 1015 Lausanne,
Switzerland}
\author{E. Mancini}
\affiliation{Fakult\"{a}t f\"{u}r Physik, University of Regensburg,
93053 Regensburg, Germany}
\author{S. Moyerman}
\affiliation{University of California, San Diego, La Jolla, CA
92093-0401, USA}
\author{E. E. Fullerton}
\affiliation{University of California, San Diego, La Jolla, CA
92093-0401, USA}
\author{R. Feidenhans'l}
\affiliation{Niels Bohr Institute, University of Copenhagen, 2100
K\o benhavn, Denmark}
\author{C. H. Back}
\affiliation{Fakult\"{a}t f\"{u}r Physik, University of Regensburg,
93053 Regensburg, Germany}
\author{C. Quitmann}
\affiliation{Swiss Light Source, Paul Scherrer Institut, 5232
Villigen, Switzerland}

\date{4 January 2012}

\begin{abstract}
We use time-resolved x-ray diffraction and magnetic optical Kerr
effect to study the laser induced antiferromagnetic to ferromagnetic
phase transition in FeRh. The structural response is given by the
nucleation of independent ferromagnetic domains ($\tau_1 \sim
30$ps). This is significantly faster than the magnetic response
($\tau_2 \sim 60$ps) given by the subsequent domain realignment.
X-ray diffraction shows that the two phases co-exist on short
time-scales and that the phase transition is limited by the speed of
sound. A nucleation model describing both the structural and
magnetic dynamics is presented.
\end{abstract}

\maketitle To date the fastest manipulation of magnetic films and
elements are induced by single fs laser pulses and include domain
switching and demagnetization on sub-ps time-scales
\cite{Beaurepaire1996,Stanciu2007,Radu2011,Kirilyuk2010}. On similar
time scales magnetization has been switched by ultrashort but strong
magnetic field pulses generated by relativistic electron bunches
\cite{Back1999} and stripline techniques \cite{Gerrits2002}, while a
third intriguing option is the manipulation of the magnetic energy
landscape by strong single cycle electric field pulses
\cite{Gamble2009}. Generation of a magnetic moment is equally
interesting but harder to achieve on an ultrafast timescale.
Ferromagnetic (FM) order can be established in ordinary FM materials
by cooling from the paramagnetic phase, but the process is limited
by heat transfer and typical timescales are nanoseconds. In this
context the antiferromagnetic (AFM) to FM phase transition in FeRh
($T_T \approx 375K$) is interesting because it can be induced by fs
laser pulses. The first order phase transition from the low
temperature AFM phase is accompanied by a $\sim$0.5 \% volume
increase, and though this has been known since 1939
\cite{Fallot1939} the physical mechanism behind the transition has
never been resolved and is still debated
\cite{Moruzzi1992,Gu2005,Sandratskii2011}.

The ultrafast transition has been studied in all-optical pump-probe
experiments using both transient reflectivity and time resolved
magneto optical Kerr effect (TR-MOKE) \cite{Thiele2004,Ju2004}.
While the reflectivity measures a combination of electronic and
structural properties, TR-MOKE measures the magnetization. In
addition x-ray magnetic circular dichroism (XMCD), which is element
specific, has been used to probe the transition \cite{Radu2010}.
This study found a gradual growth of the magnetization on a
time-scale of $\sim100$~ps. In none of these experiments was a
separate determination of the lattice dynamics possible. The
magnetization dynamics have been simulated with the
Landau-Lifshitz-Gilbert equation and a model which considers growth
of a non-homogenous magnetization proportional to the
spin-temperature followed by a slower alignment and precession of
the local magnetic moments \cite{Bergman2006}.

In order to unravel the physical processes underlying the phase
transition a prerequisite is the ability to distinguish the
contributions arising from the lattice- and magnetic- dynamics. In
this letter we report a time resolved x-ray diffraction (XRD)
experiment directly measuring the structural dynamics and providing
the evolution of the AFM and FM volume fractions with a time
resolution of $\sim200$~fs. The results are combined with TR-MOKE
measurements on the same sample and are explained using a simple
model describing the nucleation and subsequent alignment of FM
domains.

The FeRh epitaxial thin film (\textit{d} = 47 nm) was grown on
MgO(001) by co-magnetron sputtering from elemental targets. The film
is epitaxial with a (001) surface. Upon heating through the phase
transition the lattice expands 0.7 \% along the surface normal, in
contrast to the isotropic expansion of bulk samples, indicating a
strong in-plane strain.

The time-resolved XRD was performed at an x-ray energy of 7~keV
using a synchrotron slicing source (200~ph/pulse at 2~kHz and 1.2
\%bw) as probe and an 800 nm 120 fs \textit{p}-polarized laser pulse
with an incidence angle of $12^o$ as pump, resulting in a total time
resolution of 200~fs \cite{Beaud2007}. The x-ray gracing incidence
angle was either $\alpha = 0.51^o$ or $0.71^o$, in order to match
the penetration depth of the x-ray probe to either the laser pump
(penetration depth 15 nm) or to the film thickness (47 nm). Due to
the grazing angle the x-ray spot size was $0.4\times1$~mm$^2$. The
(101) Bragg reflection was recorded with a two dimensional PILATUS
100K pixel detector and rocking curves with 60 discrete images were
acquired by rotating the sample around the surface normal
($\pm2.5^o$) \cite{Mariager2009}.
\begin{figure} [t]
\includegraphics[scale=.88]{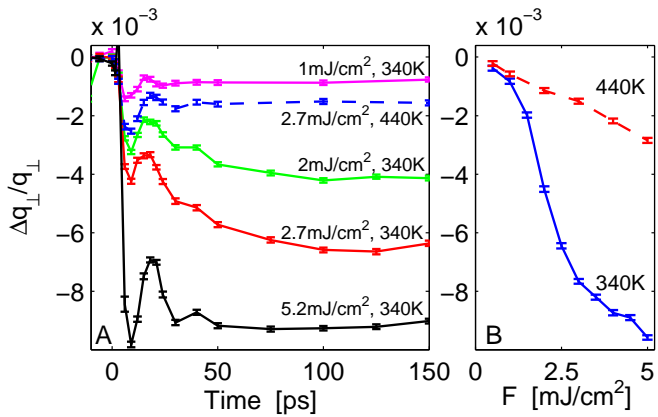}
\caption{(Color online) Shift of the center of the (101) Bragg peak
after laser excitation. (A) As a function of time delay. (B) As a
function of fluence at a fixed time delay of 145~ps for temperatures
above and below T$_T = 375K.$} \label{fig1}
\end{figure}

The TR-MOKE measurements were performed at 72~kHz as a two-color
pump-probe experiment with 200~fs cross-polarized 800~nm p-polarized
pump and 400~nm probe pulses. The probe ($58\times26$ $\mu$m$^2$)
had an incidence angle of $60^o$ with respect to the surface while
the pump was incident along the surface normal ($90^o$). We used a
longitudinal geometry with an applied in-plane external field of 0.1
T. The time dependent traces were recorded for opposite orientations
of the external field and the Kerr rotation is the difference of the
two traces \cite{Koopmans2003}.

In \figref{fig1}A we show the shift of the center of the Bragg peak
as a function of time and laser fluence. As the expansion due to the
film geometry is purely one dimensional, the shift occurs along the
surface normal ($q_\perp$). A negative peak shift corresponds to an
expansion of the lattice, and for a single unstrained phase the peak
shift is proportional to the lattice expansion, $\Delta a/a \approx
- \Delta q_{\perp}/q_{\perp}$. We first consider data obtained at a
sample temperature of \textit{T} = 340~K $<$ T$_T$. At a pump
fluence of 1~mJ/cm$^2$ the Bragg peak shifts to a new equilibrium
position through a single damped oscillation. This is the signature
of a thermally induced strain wave \cite{Thomsen1986} and the period
\textit{T}= $18.6\pm0.9$~ps (95\% confidence interval) is given by
the time it takes a strain wave to travel back and forth through the
film, \textit{T} = 2\textit{d}/\textit{v}. The resulting speed of
sound \textit{v} = 5.1~km/s is consistent with the literature
\cite{Ricodeau1972}. At higher pump fluences the strain wave is
still present but accompanied by an increased peak shift. At
intermediate (2-2.7~mJ/cm$^2$) laser fluences this extra peak shift
is slower than the strain wave, while at high laser fluences
(5.2~mJ/cm$^2$) the time scales are comparable. We attribute this
peak shift to the transformation from the AFM to the FM phase, and
since the peak shift at intermediate fluences is slower than the
strain wave, the strain wave can not be the driving force of the
transition. In \figref{fig1}B the peak shift is shown as a function
of fluence for a time delay of 145~ps. A deviation from the linear
fluence dependence given by thermal expansion, is observed above
1~mJ/cm$^2$. This threshold behavior is the signature of a
laser-induced phase transition. The threshold fluence of $\sim$1
mJ/cm$^2$ corresponds to a temperature increase of 32~K
\footnote{Calculated with reflectivity R = 0.32, penetration depth =
11~nm \cite{Rhee1995}, density 9.88~g/cm$^3$ and specific heat
capacity 465~J/kgK \cite{Annaorazov1992}.}. This matches the
difference between the sample (340~K) and the transition temperature
(T$_T$=375~K) and is consistent with the thermal nature of the laser
induced transition. As a final verification that the peak shift
arises from the AFM to FM phase transition the sample was heated to
440~K, well above T$_T$. At this temperature the peak shift is given
solely by thermal expansion. We thus unambiguously confirm the laser
induced AFM to FM phase transition.

\begin{figure} [t]
\includegraphics[scale=.7]{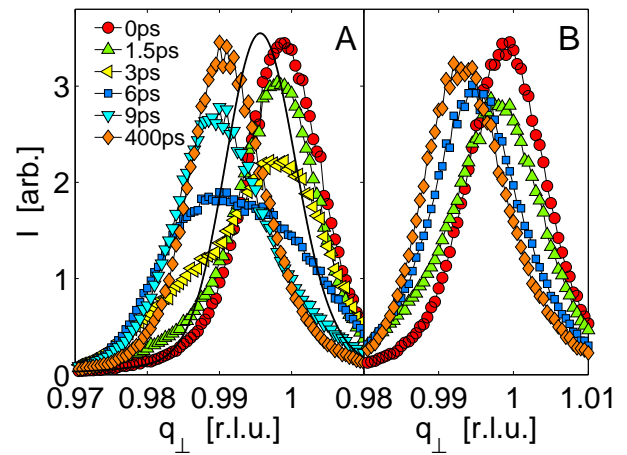}
\caption{(Color online) Bragg reflections obtained at different time
delays, (A) for a pump fluence of 5.2~mJ/cm$^2$ and (B) for a pump
fluence of 2.7~mJ/cm$^2$. The solid line is the simulated Bragg
reflection at t=6ps under the sole influence of thermal strain.}
\label{fig2}
\end{figure}

In \figref{fig2}A we further show the full Bragg peaks for several
time delays and a fixed pump fluence of 5.2~mJ/cm$^2$. At \textit{t}
= 0~ps the entire film is in the AFM phase. At \textit{t} = 400~ps
the entire film has been driven into the FM phase. The full
transformation is evident as the two peaks have the same shape. At
intermediate times the peak is a sum of the two distinct peaks
corresponding to the two phases. We thus directly confirm the
coexistence of the two structural phases at short time scales. This
is supported by the solid line which is the calculated Bragg peak at
\textit{t} = 6~ps under the sole influence of thermal strain
\cite{Thomsen1986}. It is thus evident that the transition proceeds
through the decrease of the AFM phase and increase of the FM phase,
rather than through a continuous change of the lattice constant. The
phase coexistence is a common trait of first order transitions and
has been observed statically in FeRh for both the magnetic
\cite{Radu2010} and the crystallographic structure \cite{Kim2009}.
\figref{fig2}B shows similar Bragg peaks obtained at a fixed pump
fluence of 2.7~mJ/cm$^2$. In this case the final state (\textit{t} =
400ps) consists of both AFM and FM structure.

In \figref{fig1}A it can be seen that the amplitude of the initial
strain wave does not scale with fluence above the threshold. Thus
the phase transition adds to the magnitude of the strain wave. The
fact that the phase of the FM peak is initially strained is also
seen in \figref{fig2}A. Since the stress is relaxed at the speed of
sound at least for parts of the film the underlying change in energy
landscape from AFM to FM appears faster than $\tau\sim$
\textit{d/v}. This timescale is the limit for the structural
transition.

\begin{figure} [t]
\includegraphics[scale=.68]{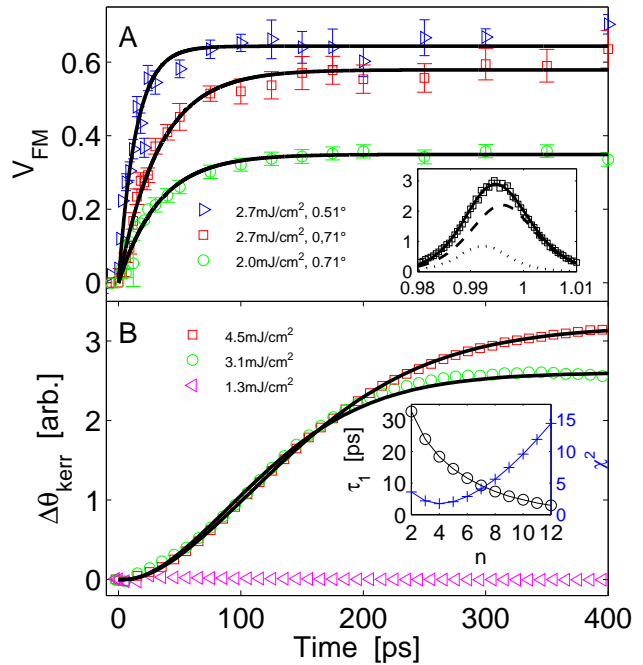}
\caption{(Color online) (A) Evolution of the volume fraction of the
expanded FM phase as a function of time delay, shown for two
different pump fluences and incidence angles. The insert illustrates
the fitting procedure (t = 6ps, 2.7~mJ/cm$^2$), showing the AFM
(dashed) and FM (dotted) fit functions, and their sum (solid). (B)
Transient Kerr rotation for comparable fluences. The inset shows the
dependence of the time constant $\tau_1$ (o) and a chi-square
estimator (+) on the model parameter n. The solid lines are fits to
the data as described in the text.} \label{fig3}
\end{figure}

To obtain a quantitative measure of the structural change due to the
phase transition we separate it from the effects of the strain wave.
In our experiment the grazing incidence geometry and the relatively
low x-ray energy limit the separation $\Delta q_{\perp}$ between the
AFM and FM peaks, but the two contributions can be systematically
extracted from the measured data by fitting it to the sum of two
symmetric functions $(a/\cosh((x-b)/c)^2)$ \footnote{To obtain the
fit the AFM Bragg peak position was fixed to the peak shift measured
at 1~mJ/cm$^2$ but scaled with fluence. At intermediate fluences,
where the timescales of the phase transition and the strain wave
differ significantly, the FM Bragg peak was restricted by an upper
limit given by the measured peak shift at 400~K. To fix the FM peak
position at $\alpha = 0.51^o$ the shift obtained from the scaled
strain wave was superimposed on the upper limit.}. One such fit is
shown in the insert in \figref{fig3}A. This way the integrated
intensities which are proportional to the scattering volumes of the
two phases can be extracted, and in \figref{fig3}A the volume
fraction of the FM phase (V$_{FM}$) is shown as a function of time
delay. Before the arrival of the laser pump the film is entirely in
the AFM phase and V$_{FM}$=0. After laser excitation the FM volume
grows to saturation within 100 - 200~ps. Increasing the fluence from
2.0 to 2.7mJ/cm$^2$ increases the transformed volume fraction of the
FM phase V$_{FM}^*$ from 32\% to 59\%. By decreasing the x-ray
penetration depth from 47~nm ($\alpha = 0.71^o$) to 15~nm ($0.51^o$)
we see that for smaller probe depths V$_{FM}$ rises significantly
faster, while the transformed volume is only increased slightly.
This implies that the nucleation of the FM phase starts at the free
surface while the final phase consists mainly of domains penetrating
the entire film depth.

To compare the change in structure to the change in magnetization we
measured the change in Kerr rotation ($\Delta\Theta_{kerr}$) on the
same sample but at 313~K, as shown in \figref{fig3}B. The higher
pump fluences applied compensate for the lower sample temperature.
As for the x-ray data we observe a threshold in laser fluence. Below
the threshold there is no change in magnetic moment (1.3~mJ/cm$^2$)
while above the threshold the Kerr rotation increases and reaches
saturation after several hundred ps (3.1 and 4.5~mJ/cm$^2$). For the
first 100~ps the dynamics appears to follow a power law, in strong
contrast to the growth of V$_{FM}$. As shown in the model below this
difference arises because XRD measures the volume of FM domains,
while TR-MOKE measures their alignment. Our TR-MOKE results are in
agreement with previous XMCD \cite{Radu2010} and TR-MOKE
\cite{Thiele2004, Ju2004} experiments, except for the previously
reported ultrafast TR-MOKE component which we do not observe. The
absence of the ultrafast response may have two reasons. First,
TR-MOKE is prone to optical artifacts in the first hundreds of fs
after excitation. Second, the absence might be due to the different
probe spot size used in the different experiments, as this
determines whether or not the magnetization is averaged over many FM
domains or just a few.

%\section{Model}
We now present a model describing the observed dynamics in terms of
nucleation and alignment of FM domains. We assume that the film is
instantaneously heated above T$_T$ and that the nucleation of the FM
phase proceeds through nucleation at many independent sites. The
rate of change of V$_{FM}$ is then proportional to V$_{AFM}$ and
described by a single time constant $\tau_1$ which may differ for
the structural and magnetic changes. Because the transition is of
first order the final state may be mixed. To account for this we
introduce the final fraction of FM phase V$_{FM}^*$. As V$_{FM}$ +
V$_{AFM}$ = 1 we then find:
\begin{equation}
\frac{dV_{FM}}{dt} = \frac{V_{AFM}-(1-V_{FM}^*)}{\tau_1} =
\frac{V_{FM}^*-V_{FM}}{\tau_1} \label{eqn1}
\end{equation}
This is essentially the Avrami model without growth of existing
domains \cite{Avrami1939} and the solution is the exponential
function $V_{FM}^*(1-\exp(-t/\tau_1))$. The depth dependence is only
included by allowing different time constants when averaging across
different probe depths. This exponential function has been used to
fit the data in \figref{fig3}A and describes the data very well.
Within error bars we find the same time constant $\tau_1 =
33\pm4$~ps for both intermediate fluences, while $\tau_1 =
14\pm3$~ps at $\alpha = 0.51^o$. The exponential growth of V$_{FM}$
is only observed when the nucleation of independent domains is
dominant while the growth of existing domains is suppressed. As the
x-ray spot size is $0.4\times1$ mm$^2$ the distance between
nucleation sites must then be less than $\sim10~\mu$m.

In order to describe the magnetization we assume that the FM domains
nucleate with the magnetization fixed into one of \textit{n}
directions and that all \textit{n} directions are equally probable.
We assume one of these directions is favored by the applied magnetic
field and define the volume fraction $V_A$ of FM phase aligned to
the applied field and the volume fraction V$_n$ not yet aligned.
These satisfy \textrm{$V_{FM}$} = \textrm{$V_n$} + \textrm{$V_A$}.
We finally assume that alignment of the magnetization occurs by
growth of the aligned FM domains at the expense of non-aligned FM
domains, as described by a product term V$_A$V$_n/\tau_2$ and a
single time-constant $\tau_2$. As this model depends on the
existence of both aligned and un-aligned FM domains it supports a
theory where short-range interactions are responsible for aligning
neighboring FM domains through domain wall motion. Given these
assumptions the time evolutions of \textrm{$V_n$} and \textrm{$V_A$}
are described by two differential equations:
\begin{equation}
\frac{dV_{n}}{dt} =
\frac{n-1}{n}\frac{dV_{FM}}{dt}-\frac{V_AV_n}{\tau_2} \label{eqn2}
\end{equation}
\begin{equation}
\frac{dV_{A}}{dt} =
\frac{1}{n}\frac{dV_{FM}}{dt}+\frac{V_AV_n}{\tau_2} \label{eqn3}
\end{equation}
The MOKE signal is proportional to the magnetization along the
preferred direction: $<$\textbf{m}$>$ $\propto$ V$_A$  -
V$_n$/(\textit{n}-1). This is valid when one of the \textit{n}-1
directions is opposite to the applied field and the magnetization of
the remaining \textit{n}-2 directions averages to zero. The
underlying assumptions of nucleation and coexistence of phases in
this model differ from the work by Bergman et al. \cite{Bergman2006}
who assumed that the local magnetization grows monotonously with
spin-temperature throughout the film.

The three differential equations are solved numerically for integer
values of n, and the result of fitting the result to the TR-MOKE
data is shown in \figref{fig3}B for \textit{n}=4, which optimizes
the fit. The agreement between experiment and fit is excellent with
$\chi^2 = 1.8$. The result \textit{n}=4 is consistent with the
in-plane magnetization expected for a cubic thin film, and has been
confirmed by static XMCD PEEM images obtained for the same film. For
the nucleation time $\tau_1=17.8\pm0.9$~ps we find, as for the XRD
data, the same value for both pump fluences within errorbars. Since
the probe depth for the laser in the TR-MOKE experiment is
$\sim$11~nm this must be compared to the $14\pm3$~ps obtained with
XRD at $\alpha = 0.51^o$. We thus conclude that the timescales of
nucleation for magnetic and structural domains are the same within
the errorbars. For the final parameter $\tau_2$ which describes the
alignment of domains we obtain $72\pm1$~ps and $57\pm1$~ps for
fluences of 4.5~mJ/cm$^2$ and 3.1~mJ/cm$^2$ respectively. Based on
the good agreement between data and model we conclude that the FM
domains initially nucleate with un-aligned moments which are
subsequently aligned. We speculate that the initial domain structure
is given by the underlying AFM phase and that the mechanism
responsible for the re-alignment process is domain wall motion.

The simplest alternative model would describe realignment as a
rotation of the total moment of a domain. The realignment term in
\eqnref{eqn2} and \eqnref{eqn3} would then be $\mp$V$_n$/$\tau_2$,
independent of V$_A$. For this model $\chi^2 \approx 30$ which is
significantly worse than $\chi^2 = 1.8$ obtained above. In addition
the fit is independent of \textit{n} and results in different
nucleation times $\tau_1$ for structure and magnetism. This would
imply that the phase transition is independent of the anisotropy and
that the magnetic domains nucleate slower than the structure. Both
conclusions appear less likely than those reached from \eqnref{eqn1}
- \eqnref{eqn3}. We thus reject the alternative explanation.

%\section{Conclusion}
In summary we have measured the structural and magnetization
dynamics of the AFM to FM phase transition in FeRh on an ultra-fast
timescale. XRD allowed us to directly observe the co-existence of
the two phases and to derive a simple model which describes the
evolution of both the structure and the magnetization. We find two
intrinsic timescales: One for the initial nucleation of FM domains
which is the same for both magnetic and structural dynamics, and a
second for the subsequent growth of FM domains aligned to the
applied magnetic field. The co-existence of FM and AFM domains is
clearly seen in the XRD data. At intermediate pump fluences the
phase transition to a large extent proceeds similarly to static
heating, while at higher fluences the structural change is limited
by the speed of sound. This speed limit on structural change in
principal allows for a significantly faster magnetic response, which
we do not observe. It thus appears that magnetic and structural
nucleation go hand in hand rather than one driving the other. While
the microscopic nature of the magnetization change has been
considered theoretically \cite{Sandratskii2011}, a more definitive
answer will require the use of spatially resolved magnetic probes in
order not to average the dynamics over many domains.

%\section{Acknowledgements}
\begin{acknowledgments}
The x-ray experiments were performed on the X05LA beam line at the
Swiss Light Source, Paul Scherrer Institut, Villigen, Switzerland
and we thank D. Grolimund and C. Borca for help. SOM thanks P.
Derlet and FP thanks G. Woltersdorf for fruitful discussions. We
acknowledge discussions with K. Sokolowski-Tinten of our results.
This work was supported by the Swiss National Foundation through
NCCR MUST, by Marie Curie Actions through the FANTOMAS Project
within the Seventh Framework Programme (FP7), by the Danish natural
science council through DANSCATT and by the Danish National Science
Foundation. Work in UCSD was partially supported by DOE-BES Award
DE-SC0003678.
\end{acknowledgments}

\bibliography{C:/referencesAllJabRef}

\begin{thebibliography}{24}
\expandafter\ifx\csname natexlab\endcsname\relax\def\natexlab#1{#1}\fi
\expandafter\ifx\csname bibnamefont\endcsname\relax
  \def\bibnamefont#1{#1}\fi
\expandafter\ifx\csname bibfnamefont\endcsname\relax
  \def\bibfnamefont#1{#1}\fi
\expandafter\ifx\csname citenamefont\endcsname\relax
  \def\citenamefont#1{#1}\fi
\expandafter\ifx\csname url\endcsname\relax
  \def\url#1{\texttt{#1}}\fi
\expandafter\ifx\csname urlprefix\endcsname\relax\def\urlprefix{URL }\fi
\providecommand{\bibinfo}[2]{#2}
\providecommand{\eprint}[2][]{\url{#2}}

\bibitem[{\citenamefont{Beaurepaire et~al.}(1996)\citenamefont{Beaurepaire,
  Merle, Daunois, and Bigot}}]{Beaurepaire1996}
\bibinfo{author}{\bibfnamefont{E.}~\bibnamefont{Beaurepaire}},
  \bibinfo{author}{\bibfnamefont{J.-C.} \bibnamefont{Merle}},
  \bibinfo{author}{\bibfnamefont{A.}~\bibnamefont{Daunois}}, \bibnamefont{and}
  \bibinfo{author}{\bibfnamefont{J.-Y.} \bibnamefont{Bigot}},
  \bibinfo{journal}{Phys. Rev. Lett.} \textbf{\bibinfo{volume}{76}},
  \bibinfo{pages}{4250} (\bibinfo{year}{1996}).

\bibitem[{\citenamefont{Stanciu et~al.}(2007)\citenamefont{Stanciu, Hansteen,
  Kimel, Kirilyuk, Tsukamoto, Itoh, and Rasing}}]{Stanciu2007}
\bibinfo{author}{\bibfnamefont{C.~D.} \bibnamefont{Stanciu}},
  \bibinfo{author}{\bibfnamefont{F.}~\bibnamefont{Hansteen}},
  \bibinfo{author}{\bibfnamefont{A.~V.} \bibnamefont{Kimel}},
  \bibinfo{author}{\bibfnamefont{A.}~\bibnamefont{Kirilyuk}},
  \bibinfo{author}{\bibfnamefont{A.}~\bibnamefont{Tsukamoto}},
  \bibinfo{author}{\bibfnamefont{A.}~\bibnamefont{Itoh}}, \bibnamefont{and}
  \bibinfo{author}{\bibfnamefont{T.}~\bibnamefont{Rasing}},
  \bibinfo{journal}{Phys. Rev. Lett.} \textbf{\bibinfo{volume}{99}},
  \bibinfo{pages}{047601} (\bibinfo{year}{2007}).

\bibitem[{\citenamefont{Radu et~al.}(2011)\citenamefont{Radu, Vahaplar, Stamm,
  Kachel, Pontius, Duerr, Ostler, Barker, Evans, Chantrell et~al.}}]{Radu2011}
\bibinfo{author}{\bibfnamefont{I.}~\bibnamefont{Radu}},
  \bibinfo{author}{\bibfnamefont{K.}~\bibnamefont{Vahaplar}},
  \bibinfo{author}{\bibfnamefont{C.}~\bibnamefont{Stamm}},
  \bibinfo{author}{\bibfnamefont{T.}~\bibnamefont{Kachel}},
  \bibinfo{author}{\bibfnamefont{N.}~\bibnamefont{Pontius}},
  \bibinfo{author}{\bibfnamefont{H.~A.} \bibnamefont{Duerr}},
  \bibinfo{author}{\bibfnamefont{T.~A.} \bibnamefont{Ostler}},
  \bibinfo{author}{\bibfnamefont{J.}~\bibnamefont{Barker}},
  \bibinfo{author}{\bibfnamefont{R.~F.~L.} \bibnamefont{Evans}},
  \bibinfo{author}{\bibfnamefont{R.~W.} \bibnamefont{Chantrell}},
  \bibnamefont{et~al.}, \bibinfo{journal}{Nature}
  \textbf{\bibinfo{volume}{472}}, \bibinfo{pages}{205} (\bibinfo{year}{2011}).

\bibitem[{\citenamefont{Kirilyuk et~al.}(2010)\citenamefont{Kirilyuk, Kimel,
  and Rasing}}]{Kirilyuk2010}
\bibinfo{author}{\bibfnamefont{A.}~\bibnamefont{Kirilyuk}},
  \bibinfo{author}{\bibfnamefont{A.~V.} \bibnamefont{Kimel}}, \bibnamefont{and}
  \bibinfo{author}{\bibfnamefont{T.}~\bibnamefont{Rasing}},
  \bibinfo{journal}{{Rev. Mod. Phys.}} \textbf{\bibinfo{volume}{82}},
  \bibinfo{pages}{2731} (\bibinfo{year}{2010}).

\bibitem[{\citenamefont{Back et~al.}(1999)\citenamefont{Back, Allenspach,
  Weber, Parkin, Weller, Garwin, and Siegmann}}]{Back1999}
\bibinfo{author}{\bibfnamefont{C.~H.} \bibnamefont{Back}},
  \bibinfo{author}{\bibfnamefont{R.}~\bibnamefont{Allenspach}},
  \bibinfo{author}{\bibfnamefont{W.}~\bibnamefont{Weber}},
  \bibinfo{author}{\bibfnamefont{S.~S.~P.} \bibnamefont{Parkin}},
  \bibinfo{author}{\bibfnamefont{D.}~\bibnamefont{Weller}},
  \bibinfo{author}{\bibfnamefont{E.~L.} \bibnamefont{Garwin}},
  \bibnamefont{and} \bibinfo{author}{\bibfnamefont{H.~C.}
  \bibnamefont{Siegmann}}, \bibinfo{journal}{Science}
  \textbf{\bibinfo{volume}{285}}, \bibinfo{pages}{864} (\bibinfo{year}{1999}).

\bibitem[{\citenamefont{Gerrits et~al.}(2002)\citenamefont{Gerrits, van~den
  Berg, Hohlfeld, Bar, and Rasing}}]{Gerrits2002}
\bibinfo{author}{\bibfnamefont{T.}~\bibnamefont{Gerrits}},
  \bibinfo{author}{\bibfnamefont{H.}~\bibnamefont{van~den Berg}},
  \bibinfo{author}{\bibfnamefont{J.}~\bibnamefont{Hohlfeld}},
  \bibinfo{author}{\bibfnamefont{L.}~\bibnamefont{Bar}}, \bibnamefont{and}
  \bibinfo{author}{\bibfnamefont{T.}~\bibnamefont{Rasing}},
  \bibinfo{journal}{Nature} \textbf{\bibinfo{volume}{418}},
  \bibinfo{pages}{509} (\bibinfo{year}{2002}).

\bibitem[{\citenamefont{Gamble et~al.}(2009)\citenamefont{Gamble, Burkhardt,
  Kashuba, Allenspach, Parkin, Siegmann, and Stohr}}]{Gamble2009}
\bibinfo{author}{\bibfnamefont{S.~J.} \bibnamefont{Gamble}},
  \bibinfo{author}{\bibfnamefont{M.~H.} \bibnamefont{Burkhardt}},
  \bibinfo{author}{\bibfnamefont{A.}~\bibnamefont{Kashuba}},
  \bibinfo{author}{\bibfnamefont{R.}~\bibnamefont{Allenspach}},
  \bibinfo{author}{\bibfnamefont{S.~S.~P.} \bibnamefont{Parkin}},
  \bibinfo{author}{\bibfnamefont{H.~C.} \bibnamefont{Siegmann}},
  \bibnamefont{and} \bibinfo{author}{\bibfnamefont{J.}~\bibnamefont{Stohr}},
  \bibinfo{journal}{Phys. Rev. Lett.} \textbf{\bibinfo{volume}{102}},
  \bibinfo{pages}{217201} (\bibinfo{year}{2009}).

\bibitem[{\citenamefont{Fallot and Hocart}(1939)}]{Fallot1939}
\bibinfo{author}{\bibfnamefont{M.}~\bibnamefont{Fallot}} \bibnamefont{and}
  \bibinfo{author}{\bibfnamefont{R.}~\bibnamefont{Hocart}},
  \bibinfo{journal}{{Rev. Sci.}} \textbf{\bibinfo{volume}{77}},
  \bibinfo{pages}{498} (\bibinfo{year}{1939}).

\bibitem[{\citenamefont{Moruzzi and Marcus}(1992)}]{Moruzzi1992}
\bibinfo{author}{\bibfnamefont{V.~L.} \bibnamefont{Moruzzi}} \bibnamefont{and}
  \bibinfo{author}{\bibfnamefont{P.~M.} \bibnamefont{Marcus}},
  \bibinfo{journal}{Phys. Rev. B} \textbf{\bibinfo{volume}{46}},
  \bibinfo{pages}{2864} (\bibinfo{year}{1992}).

\bibitem[{\citenamefont{Gu and Antropov}(2005)}]{Gu2005}
\bibinfo{author}{\bibfnamefont{R.~Y.} \bibnamefont{Gu}} \bibnamefont{and}
  \bibinfo{author}{\bibfnamefont{V.~P.} \bibnamefont{Antropov}},
  \bibinfo{journal}{Phys. Rev. B} \textbf{\bibinfo{volume}{72}},
  \bibinfo{pages}{012403} (\bibinfo{year}{2005}).

\bibitem[{\citenamefont{Sandratskii and Mavropoulos}(2011)}]{Sandratskii2011}
\bibinfo{author}{\bibfnamefont{L.~M.} \bibnamefont{Sandratskii}}
  \bibnamefont{and}
  \bibinfo{author}{\bibfnamefont{P.}~\bibnamefont{Mavropoulos}},
  \bibinfo{journal}{Phys. Rev. B} \textbf{\bibinfo{volume}{83}},
  \bibinfo{pages}{174408} (\bibinfo{year}{2011}).

\bibitem[{\citenamefont{Thiele et~al.}(2004)\citenamefont{Thiele, Buess, and
  Back}}]{Thiele2004}
\bibinfo{author}{\bibfnamefont{J.~U.} \bibnamefont{Thiele}},
  \bibinfo{author}{\bibfnamefont{M.}~\bibnamefont{Buess}}, \bibnamefont{and}
  \bibinfo{author}{\bibfnamefont{C.~H.} \bibnamefont{Back}},
  \bibinfo{journal}{{Appl. Phys. Lett.}} \textbf{\bibinfo{volume}{85}},
  \bibinfo{pages}{2857} (\bibinfo{year}{2004}).

\bibitem[{\citenamefont{Ju et~al.}(2004)\citenamefont{Ju, Hohlfeld, Bergman,
  van~de Veerdonk, Mryasov, Kim, Wu, Weller, and Koopmans}}]{Ju2004}
\bibinfo{author}{\bibfnamefont{G.~P.} \bibnamefont{Ju}},
  \bibinfo{author}{\bibfnamefont{J.}~\bibnamefont{Hohlfeld}},
  \bibinfo{author}{\bibfnamefont{B.}~\bibnamefont{Bergman}},
  \bibinfo{author}{\bibfnamefont{R.~J.~M.} \bibnamefont{van~de Veerdonk}},
  \bibinfo{author}{\bibfnamefont{O.~N.} \bibnamefont{Mryasov}},
  \bibinfo{author}{\bibfnamefont{J.~Y.} \bibnamefont{Kim}},
  \bibinfo{author}{\bibfnamefont{X.~W.} \bibnamefont{Wu}},
  \bibinfo{author}{\bibfnamefont{D.}~\bibnamefont{Weller}}, \bibnamefont{and}
  \bibinfo{author}{\bibfnamefont{B.}~\bibnamefont{Koopmans}},
  \bibinfo{journal}{Phys. Rev. Lett.} \textbf{\bibinfo{volume}{93}},
  \bibinfo{pages}{197403} (\bibinfo{year}{2004}).

\bibitem[{\citenamefont{Radu et~al.}(2010)\citenamefont{Radu, Stamm, Pontius,
  Kachel, Ramm, Thiele, D\"urr, and Back}}]{Radu2010}
\bibinfo{author}{\bibfnamefont{I.}~\bibnamefont{Radu}},
  \bibinfo{author}{\bibfnamefont{C.}~\bibnamefont{Stamm}},
  \bibinfo{author}{\bibfnamefont{N.}~\bibnamefont{Pontius}},
  \bibinfo{author}{\bibfnamefont{T.}~\bibnamefont{Kachel}},
  \bibinfo{author}{\bibfnamefont{P.}~\bibnamefont{Ramm}},
  \bibinfo{author}{\bibfnamefont{J.-U.} \bibnamefont{Thiele}},
  \bibinfo{author}{\bibfnamefont{H.~A.} \bibnamefont{D\"urr}},
  \bibnamefont{and} \bibinfo{author}{\bibfnamefont{C.~H.} \bibnamefont{Back}},
  \bibinfo{journal}{Phys. Rev. B} \textbf{\bibinfo{volume}{81}},
  \bibinfo{pages}{104415} (\bibinfo{year}{2010}).

\bibitem[{\citenamefont{Bergman et~al.}(2006)\citenamefont{Bergman, Ju,
  Hohlfeld, van~de Veerdonk, Kim, Wu, Weller, and Koopmans}}]{Bergman2006}
\bibinfo{author}{\bibfnamefont{B.}~\bibnamefont{Bergman}},
  \bibinfo{author}{\bibfnamefont{G.}~\bibnamefont{Ju}},
  \bibinfo{author}{\bibfnamefont{J.}~\bibnamefont{Hohlfeld}},
  \bibinfo{author}{\bibfnamefont{R.~J.~M.} \bibnamefont{van~de Veerdonk}},
  \bibinfo{author}{\bibfnamefont{J.-Y.} \bibnamefont{Kim}},
  \bibinfo{author}{\bibfnamefont{X.}~\bibnamefont{Wu}},
  \bibinfo{author}{\bibfnamefont{D.}~\bibnamefont{Weller}}, \bibnamefont{and}
  \bibinfo{author}{\bibfnamefont{B.}~\bibnamefont{Koopmans}},
  \bibinfo{journal}{Phys. Rev. B} \textbf{\bibinfo{volume}{73}},
  \bibinfo{pages}{060407} (\bibinfo{year}{2006}).

\bibitem[{\citenamefont{Beaud et~al.}(2007)\citenamefont{Beaud, Johnson,
  Streun, Abela, Abramsohn, Grolimund, Krasniqi, Schmidt, Schlott, and
  Ingold}}]{Beaud2007}
\bibinfo{author}{\bibfnamefont{P.}~\bibnamefont{Beaud}},
  \bibinfo{author}{\bibfnamefont{S.~L.} \bibnamefont{Johnson}},
  \bibinfo{author}{\bibfnamefont{A.}~\bibnamefont{Streun}},
  \bibinfo{author}{\bibfnamefont{R.}~\bibnamefont{Abela}},
  \bibinfo{author}{\bibfnamefont{D.}~\bibnamefont{Abramsohn}},
  \bibinfo{author}{\bibfnamefont{D.}~\bibnamefont{Grolimund}},
  \bibinfo{author}{\bibfnamefont{F.}~\bibnamefont{Krasniqi}},
  \bibinfo{author}{\bibfnamefont{T.}~\bibnamefont{Schmidt}},
  \bibinfo{author}{\bibfnamefont{V.}~\bibnamefont{Schlott}}, \bibnamefont{and}
  \bibinfo{author}{\bibfnamefont{G.}~\bibnamefont{Ingold}},
  \bibinfo{journal}{Phys. Rev. Lett.} \textbf{\bibinfo{volume}{99}},
  \bibinfo{pages}{174801} (\bibinfo{year}{2007}).

\bibitem[{\citenamefont{Mariager et~al.}(2009)\citenamefont{Mariager,
  Lauridsen, Dohn, Bovet, Sorensen, Schlepuetz, Willmott, and
  Feidenhans'l}}]{Mariager2009}
\bibinfo{author}{\bibfnamefont{S.~O.} \bibnamefont{Mariager}},
  \bibinfo{author}{\bibfnamefont{S.~L.} \bibnamefont{Lauridsen}},
  \bibinfo{author}{\bibfnamefont{A.}~\bibnamefont{Dohn}},
  \bibinfo{author}{\bibfnamefont{N.}~\bibnamefont{Bovet}},
  \bibinfo{author}{\bibfnamefont{C.~B.} \bibnamefont{Sorensen}},
  \bibinfo{author}{\bibfnamefont{C.~M.} \bibnamefont{Schlepuetz}},
  \bibinfo{author}{\bibfnamefont{P.~R.} \bibnamefont{Willmott}},
  \bibnamefont{and}
  \bibinfo{author}{\bibfnamefont{R.}~\bibnamefont{Feidenhans'l}},
  \bibinfo{journal}{{J. Appl. Cryst.}} \textbf{\bibinfo{volume}{42}},
  \bibinfo{pages}{369} (\bibinfo{year}{2009}).

\bibitem[{\citenamefont{Koopmans}(2003)}]{Koopmans2003}
\bibinfo{author}{\bibfnamefont{B.}~\bibnamefont{Koopmans}}, in
  \emph{\bibinfo{booktitle}{Spin dynamics in confined magnetic structures II}}
  (\bibinfo{publisher}{Springer}, \bibinfo{year}{2003}).

\bibitem[{\citenamefont{Thomsen et~al.}(1986)\citenamefont{Thomsen, Grahn,
  Maris, and Tauc}}]{Thomsen1986}
\bibinfo{author}{\bibfnamefont{C.}~\bibnamefont{Thomsen}},
  \bibinfo{author}{\bibfnamefont{H.~T.} \bibnamefont{Grahn}},
  \bibinfo{author}{\bibfnamefont{H.~J.} \bibnamefont{Maris}}, \bibnamefont{and}
  \bibinfo{author}{\bibfnamefont{J.}~\bibnamefont{Tauc}},
  \bibinfo{journal}{Phys. Rev. B} \textbf{\bibinfo{volume}{34}},
  \bibinfo{pages}{4129} (\bibinfo{year}{1986}).

\bibitem[{\citenamefont{Ricodeau and Melville}({1972})}]{Ricodeau1972}
\bibinfo{author}{\bibfnamefont{J.~A.} \bibnamefont{Ricodeau}} \bibnamefont{and}
  \bibinfo{author}{\bibfnamefont{D.}~\bibnamefont{Melville}},
  \bibinfo{journal}{{J. Phys. F - Metal phys.}} \textbf{\bibinfo{volume}{{2}}},
  \bibinfo{pages}{{337}} (\bibinfo{year}{{1972}}).

\bibitem[{\citenamefont{Kim et~al.}(2009)\citenamefont{Kim, Ryan, Ding, Lewis,
  Ali, Kinane, Hickey, Marrows, and Arena}}]{Kim2009}
\bibinfo{author}{\bibfnamefont{J.~W.} \bibnamefont{Kim}},
  \bibinfo{author}{\bibfnamefont{P.~J.} \bibnamefont{Ryan}},
  \bibinfo{author}{\bibfnamefont{Y.}~\bibnamefont{Ding}},
  \bibinfo{author}{\bibfnamefont{L.~H.} \bibnamefont{Lewis}},
  \bibinfo{author}{\bibfnamefont{M.}~\bibnamefont{Ali}},
  \bibinfo{author}{\bibfnamefont{C.~J.} \bibnamefont{Kinane}},
  \bibinfo{author}{\bibfnamefont{B.~J.} \bibnamefont{Hickey}},
  \bibinfo{author}{\bibfnamefont{C.~H.} \bibnamefont{Marrows}},
  \bibnamefont{and} \bibinfo{author}{\bibfnamefont{D.~A.} \bibnamefont{Arena}},
  \bibinfo{journal}{Appl. Phys. Lett.} \textbf{\bibinfo{volume}{95}},
  \bibinfo{pages}{222515} (\bibinfo{year}{2009}).

\bibitem[{\citenamefont{Avrami}(1939)}]{Avrami1939}
\bibinfo{author}{\bibfnamefont{M.}~\bibnamefont{Avrami}}, \bibinfo{journal}{J.
  Chem. Phys.} \textbf{\bibinfo{volume}{7}}, \bibinfo{pages}{1103}
  (\bibinfo{year}{1939}).

\bibitem[{\citenamefont{Rhee and Lynch}(1995)}]{Rhee1995}
\bibinfo{author}{\bibfnamefont{J.~Y.} \bibnamefont{Rhee}} \bibnamefont{and}
  \bibinfo{author}{\bibfnamefont{D.~W.} \bibnamefont{Lynch}},
  \bibinfo{journal}{Phys. Rev. B} \textbf{\bibinfo{volume}{51}},
  \bibinfo{pages}{1926} (\bibinfo{year}{1995}).

\bibitem[{\citenamefont{Annaorazov et~al.}(1992)\citenamefont{Annaorazov,
  Asatryan, Myalikgulyev, Nikitin, Tishin, and Tyurin}}]{Annaorazov1992}
\bibinfo{author}{\bibfnamefont{M.~P.} \bibnamefont{Annaorazov}},
  \bibinfo{author}{\bibfnamefont{K.~A.} \bibnamefont{Asatryan}},
  \bibinfo{author}{\bibfnamefont{G.}~\bibnamefont{Myalikgulyev}},
  \bibinfo{author}{\bibfnamefont{S.~A.} \bibnamefont{Nikitin}},
  \bibinfo{author}{\bibfnamefont{A.~M.} \bibnamefont{Tishin}},
  \bibnamefont{and} \bibinfo{author}{\bibfnamefont{A.~L.}
  \bibnamefont{Tyurin}}, \bibinfo{journal}{Cryogenics}
  \textbf{\bibinfo{volume}{32}}, \bibinfo{pages}{867} (\bibinfo{year}{1992}).

\end{thebibliography}

\end{document}